\documentclass[a4paper,11pt]{article}
\pdfoutput=1 % if your are submitting a pdflatex (i.e. if you have
\usepackage{jinstpub} % for details on the use of the package, please
\usepackage{subfig}
\usepackage{wrapfig}

\title{\boldmath Development of gaseous particle detectors based on semiconductive plate electrodes}

\author[a,1]{A.~Rocchi,%
\note{Corresponding author.}}

\author[b,1]{R.~Cardarelli,}
\author[a]{and G.~Aielli,}
\author[a]{E.~Alunno Camelia,}
\author[a]{S.~Bruno,}
\author[a]{A.~Caltabiano,}
\author[a]{P.~Camarri,}
\author[a]{A.~Di Ciaccio,}
\author[b]{B.~Liberti,}
\author[b]{L.~Massa,}
\author[a]{L.~Pizzimento,}

\affiliation[a]{University of Rome Tor Vergata,\\
Via della Ricerca Scientifica 1, Rome, Italy}
\affiliation[b]{INFN Rome Tor Vergata,\\
Via della Ricerca Scientifica 1, Rome, Italy}

\emailAdd{alessandro.rocchi@roma2.infn.it}

\abstract{A new particle detector with sub-nanosecond time resolution capable of working in high-rate environment (rate capability of the order of $\rm MHz/\rm cm^2$) is under developmnet. Semi-conductive electrodes with resistivity $\rho$ up to $10^8\;\Omega\cdot\rm cm$ have been used to improve the RPC \cite{a} \cite{b} rate capability.
In this paper efficiency and time resolution of three different detector structures are presented.}


\collaboration[c]{}

\proceeding{14$^{\text{th}}$ Workshop on Resistive Plate Chambers and Related Detectors\\
  19-23 February 2018\\
  Puerto Vallarta, Jalisco State, MEXICO}

\begin{document}
\maketitle
\flushbottom

\section{Introduction}

In the last fourty years, Resistive Plate Chamber detectors achieved many progress. The discovery of the avalanche regime allowed the integrated charge to decrease as well as the rate capability and time resolution to increase \cite{i}. Systematic studies on the gas mixture components hilighted that electornegative compounds, like sulfur hexafluoride $SF_{6}$, move the avalanche to streamer transition to higher field value, making the  avalanche mode useful\cite{c}\cite{d}. 
This development paved the way to the application of the RPC in the collider physics experiments.
Just at early 00s, RPC detectors with $1\;\rm kHz/cm^2$ of rate capability were installed in ATLAS experiment and contributed to the Higgs boson discovery in 2013. 
The main concept on which the research has been based until now consists in the transfer of the detector gain from the gas gap to the electronics front-end, without changing the electrode material \cite{g}. Following this idea, a new generation of RPC detectors with $10\;\rm kHz/cm^2$ of rate capability was developed and will be installed in ATLAS experiment in 2019 \cite{e} \cite{f}. 
Due to the increase of collider luminosity it would be extremely useful to develop a new detector similar to the RPC as far as time resolution and radiation hardness are concerned, but with a rate capability of the order of $\rm MHz/\rm cm^2$. To rich this goal it is needed to tacking into account the replacement of the electrode material with one of lower resistivity as well as the electronics improvement. Semi-insulating Gallium Arsenide has been identified as possible upgrade of the standard bakelite electrode, resistivity of the order of $\rho\sim 10^{11}\;\rm\Omega\cdot cm$, because of its lower resistivity $\rho\sim 10^8\;\rm\Omega\cdot cm$.

\section{Increasing the Rate Capability}
\begin{figure}[h]
\centering
{\includegraphics[width=0.8\textwidth]{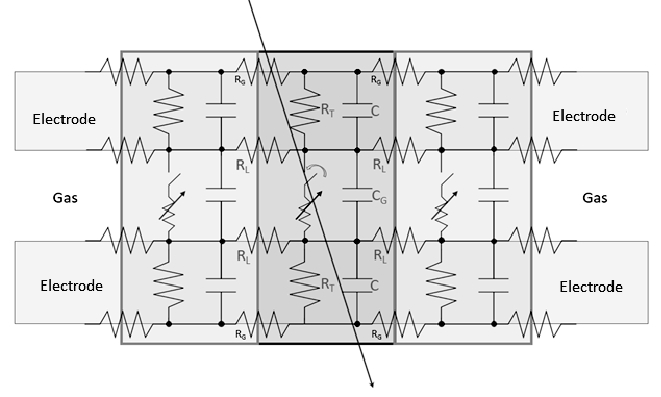}} \\
\caption{Semplified equivalent circuit superimposed on an RPC detector sketch.}
\label{fig:RC}
\end{figure}

From a microscopic point of view, an RPC detector can be described by a finite element model like a set of unit cells interconnected according to the diagram in fig. \ref{fig:RC}. A unit cell is characterized by the gas capacitance $C_G$ , the electrodes capacitances $C$ and by $R_T$  and $R_L$  resistances, which represent respectively the electrode resistance in the normal and parallel directions with respect to the electrode surface. $R_g$ is the resistance of the graphite layer which distributes the high voltage $V$ on the electrode surface.

 If the high voltage across the gas gap $V_{gas}$ is high enough (applied electic field $E$ greater than $5\;\rm kV/\rm mm$ for $95\%$/$4.5\%$/$0.5\%$ of C$_{2}$H$_{2}$F$_{4}$ - iC$_{4}$H$_{10}$ - SF$_{6}$  gas mixture), a ionizing particle crossing a unit cell could trigger the gas discharge. This process can be seen as the closure of the switch in the equivalent circuit of the unit cell involved in the discharge. The longitudinal resistance $R_L$, much greater than the graphyte resistence $R_g$, restricts the transfer of energy from adjacent cells. Increasing $R_L$ involves the unit cells density increases. The resistance $R_T$, instead, contributes with the gas capacitance $C_g$, to the dead time constant. To improve the rate capability performances is needed to reduce the dead time and to increase the unit cells density, therefore change the $R_T/R_L$ ratio working on the electrode material and thikness.
 
From a macroscopic point of view, when a charged-particle flux $\Phi$ crosses an RPC detector, the simultaneous ignition of many unit cells occurs  and the cumulative effect causes a mean voltage drop on the electrodes. 
Expressing the electrode resistance as a function of the electrode resistivity $\rho$ and the thickness $d$, 
the effective voltage on the gas gap can be written as in Eq. \ref{eq1} with $\langle Q\rangle$ representing the mean charge involved in a single process and $\Phi_{eff}$ the mean number of processes occurring in the detector per unit time and surface. 
\begin{equation}
 \label{eq1}
 V_{gas}=V-2\rho d \langle Q \rangle \Phi_{eff}
\end{equation}

\noindent The effective flux $\Phi_{eff}$ contains the contribution of both the spontaneus noise and the ionizing-particles events, the latter being a fraction $\epsilon\Phi$ of the particle flux, where $\epsilon$ is the efficiency and it depends on the voltage across the gas gap $V_{gas}$. For $\epsilon\sim 1$ and $\Phi$ much greater then the noise flux, then $\Phi_{eff}\sim\Phi$. To prevent the detector from losing efficiency as $\Phi$ rises, it is necessary to minimize the mean voltage drop on the electrodes in such a way to fix the $V_{gas}$ value. For this purpose two strategies have been combined during the test:

\begin{itemize}
\item reduction of the average charge $\langle Q \rangle$ using a charge amplifier with high signal-to-noise ratio \cite{g};

\item replacement of the standard insulating electrodes with Semi-Insulating electrodes with lower resistivity, $\rho$ and thickness $d$ and higher electorns mobility.
\end{itemize}

\section{Prototypes}
Three detectors have been built with different gas gap and electrode material. All the electrodes were $400\;\mu\rm m$ thick and were spaced with a PET circular crown. 
The first prototype was made of two SI-GaAs electrodes with a resistivity $\rho$ of the order of $10^8\; \Omega\cdot \rm cm$. The spacer was $1\;\rm mm$ thick and the gas gap was filled with a gas mixture of  C$_{2}$H$_{2}$F$_{4}$ - iC$_{4}$H$_{10}$ - SF$_{6}$ ($95\%$/$4.5\%$/$0.5\%$). 
 
The second prototype had $1.5\;\rm mm$ gas gap, one Silicon electrode with a resistivity $\rho$ of the order of $10^4\; \Omega\cdot \rm cm$ and one SI-GaAs electrode with a resistivity $\rho$ of the order of $10^8\; \Omega\cdot \rm cm$. In this case the gas gap was filled with a gas mixture of iC$_{4}$H$_{10}$ - Ar ($60\%$/$40\%$).

The third prototype had $1.3\;\rm mm$ gas gap, both SI-GaAs electrode with a resistivity $\rho$ of the order of $10^8\; \Omega\cdot \rm cm$ and was filled with a gas mixture of C$_{2}$H$_{2}$F$_{4}$ - iC$_{4}$H$_{10}$ - SF$_{6}$ ($95\%$/$4.5\%$/$0.5\%$).

\begin{figure}[h]
\centering
{\includegraphics[width=0.7\textwidth]{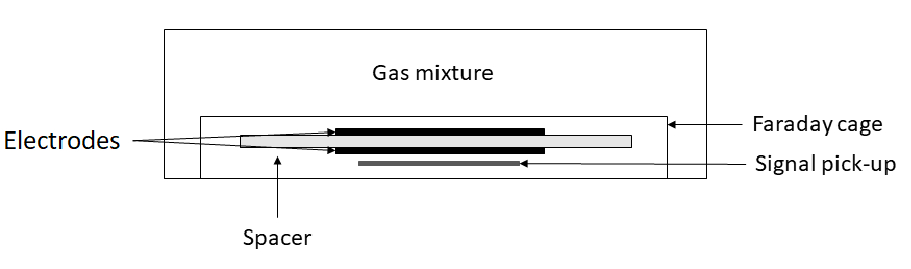}} \\
\caption{Prototype sketch.}
\label{fig:Prototype_sketch}
\end{figure}

In all the detectors, the electronic signal was read with a pad placed under the low-voltage electrode (see fig. \ref{fig:Prototype_sketch}). 
The low noise conditions of the third detector allow the study of the signal produced by the ions drift. This signal, because of the low ions drift speed, is caracterized by a low amplitude and hundreds microseconds duration. The ionic signal was read on a $50\Omega$ resistance between the low voltage electrode and the ground reference.

The first two detectors were placed in series with a $100\;\rm M\Omega$ resistance in order to avoid that the power dissipated in the electrode due to any anomalous electric discharge may damage the crystal. The third detector, because of its stability don't need the quenchig resistance.
The operational parameters of the charge amplifier used for the FE electronics are listed in Tab. \ref{table:I}.

\begin{table}[ht]
\caption{} 
\centering 
\begin{tabular}{c c}\hline
Voltage supply&3-5 Volt\\ \hline
Sensitivity&2-4 mV/fC\\ \hline
Noise(up to 20pF input capacitance)&1000 $e^{-}$ RMS\\ \hline
Input impedance&100-50 Ohm\\ \hline
B.W.&10-100 MHz\\ \hline
Power consumption&10mW/ch\\ \hline
Radiation hardness&1Mrad, $10^{13}$ n  $\rm cm^{-2}$\\ \hline
\end{tabular}
\label{table:I}
\end{table}
\newpage
\section{Experimental test}
\begin{figure}[h]
\centering
\subfloat[][\small{Time difference with respect to the trigger detector corrected for the time-walk effect ($HV = 5630\; \rm V$, prototype 1);}]
{\includegraphics[trim=0.3cm 0.1cm 0.3cm 0.3cm, clip=true, width=0.46\textwidth]{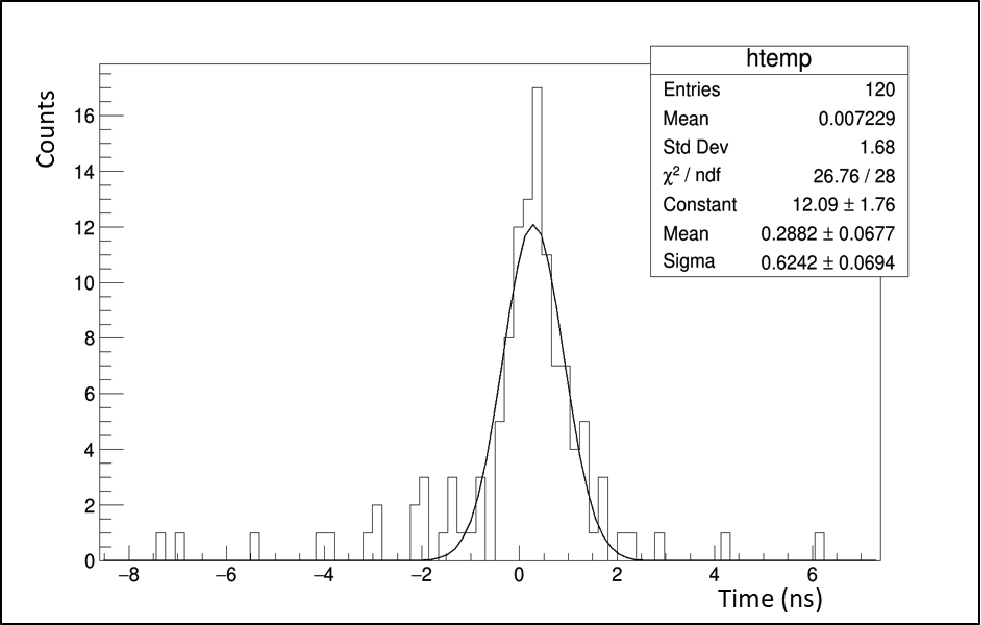}} \quad
\subfloat[][\small{'Efficiency times acceptance' ($1\;\rm mm$ gas gap, prototype 1);}]
{\includegraphics[trim=0.3cm 0.1cm 0.3cm 0.3cm, clip=true,width=0.46\textwidth]{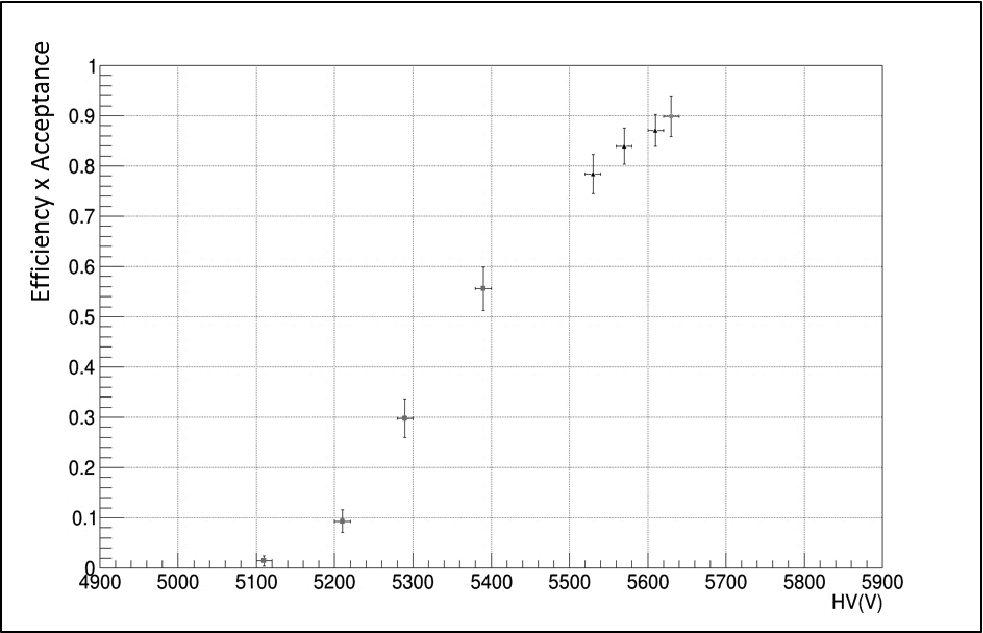}} \\
\subfloat[][\small{Pulse samples, prototype 1}]
{\includegraphics[trim=0.3cm 0.25cm 0.3cm 0.3cm, clip=true,width=0.45\textwidth]{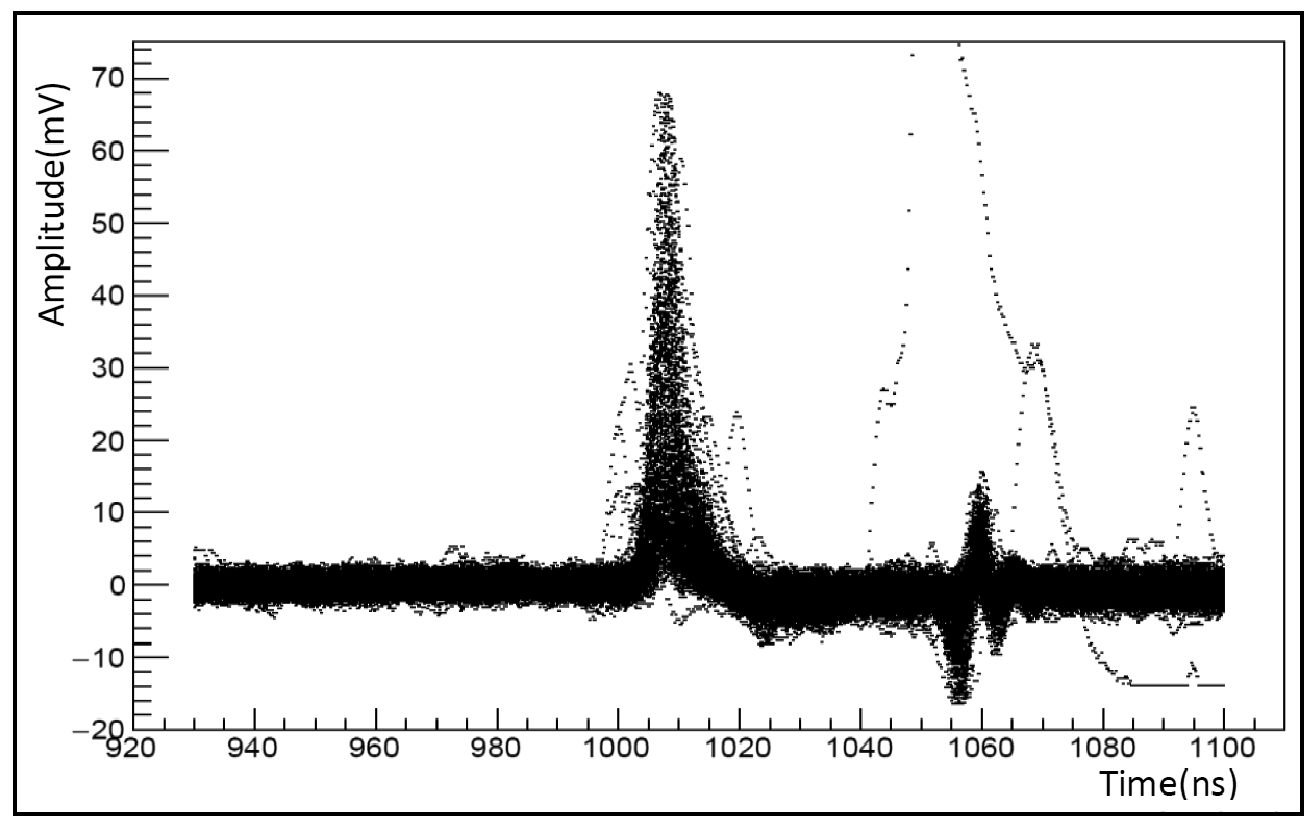}} \\
\caption{Prototype 1 results.}
\label{fig:Prototype1}
\end{figure}

The characterization of the first prototype was carried out at the Beam Test Facility (BTF) of the National Laboratories of Frascati with a beam of $450\;\rm MeV$ electrons. The average multiplicity of particles per bunch was fixed at $0.3$ for the whole duration of the test. Two silicon detectors \cite{h} optimized for time-of-flight measurements have been used as the trigger reference. The trigger-time resolution has been measured during the test, resulting in $(180\pm4)\; \rm ps$.
A summary of the results is shown in fig. \ref{fig:Prototype1}. The time reslution has been evaluated by measuring the jitter of the electron time of flight with respect to one silicon detector, corrected for the time-walk effect. A jitter of $(590\pm90)\; \rm ps$ was measured. The 'efficiency times acceptance' curve show a knee at about $5600\; \rm V$. 
 
\begin{figure}[h]
\centering
\subfloat[][\small{Time difference with respect to the trigger detector corrected for the time-walk effect (prototipe 2);}]
{\includegraphics[trim=0.cm 0.cm 0.1cm 0.cm, clip=true,width=0.45\textwidth]{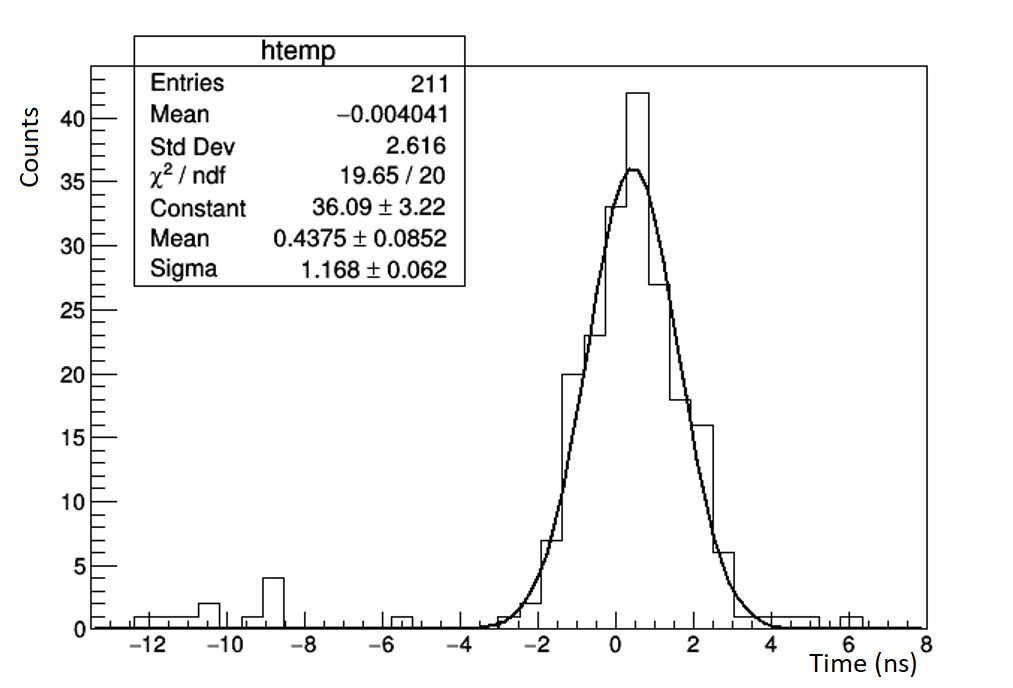}} \quad
\subfloat[][\small{'Efficiency times acceptance' ($1.5\;\rm mm$ gas gap, prototype 2);}]
{\includegraphics[trim=0.cm 0.cm 0.1cm 0.cm, clip=true,width=0.45\textwidth]{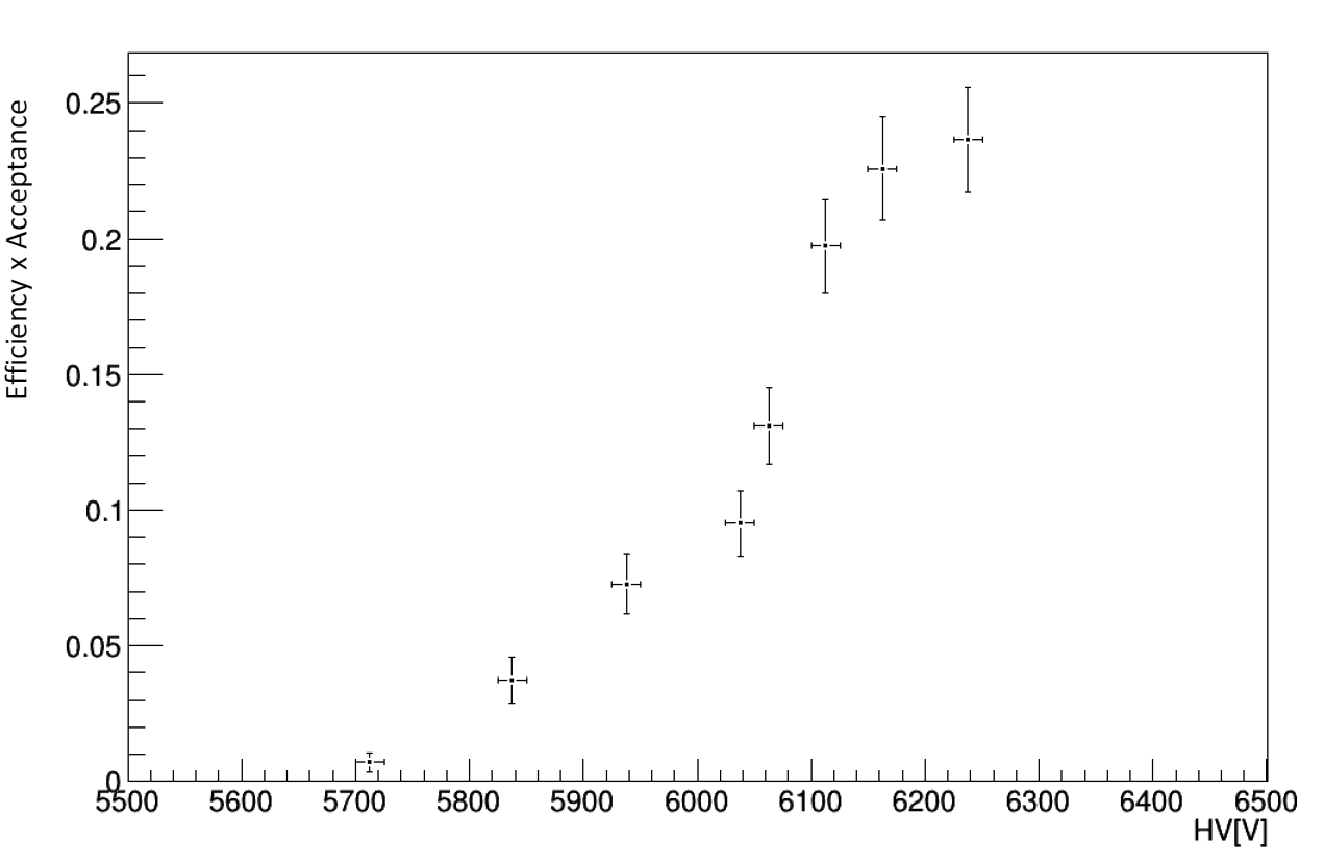}} \\
\subfloat[][\small{Pulse samples, prototype 2}]
{\includegraphics[trim=0.cm 0.cm 0.1cm 0.cm, clip=true,width=0.45\textwidth]{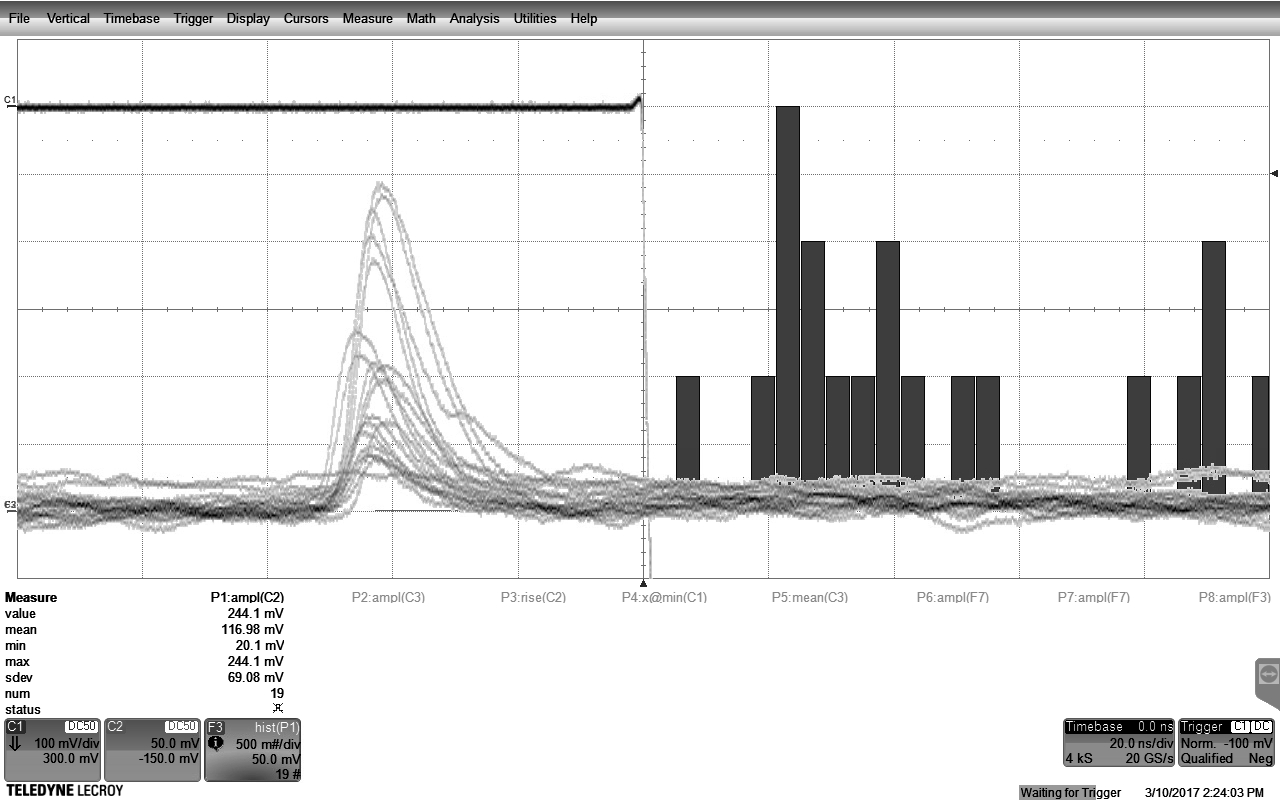}} \quad
\subfloat[][\small{Random counting rate as a function of high voltage (surface$\sim3\;\rm cm^2$).}]
{\includegraphics[trim=0.cm 0.cm 0.1cm 0.cm, clip=true,width=0.45\textwidth]{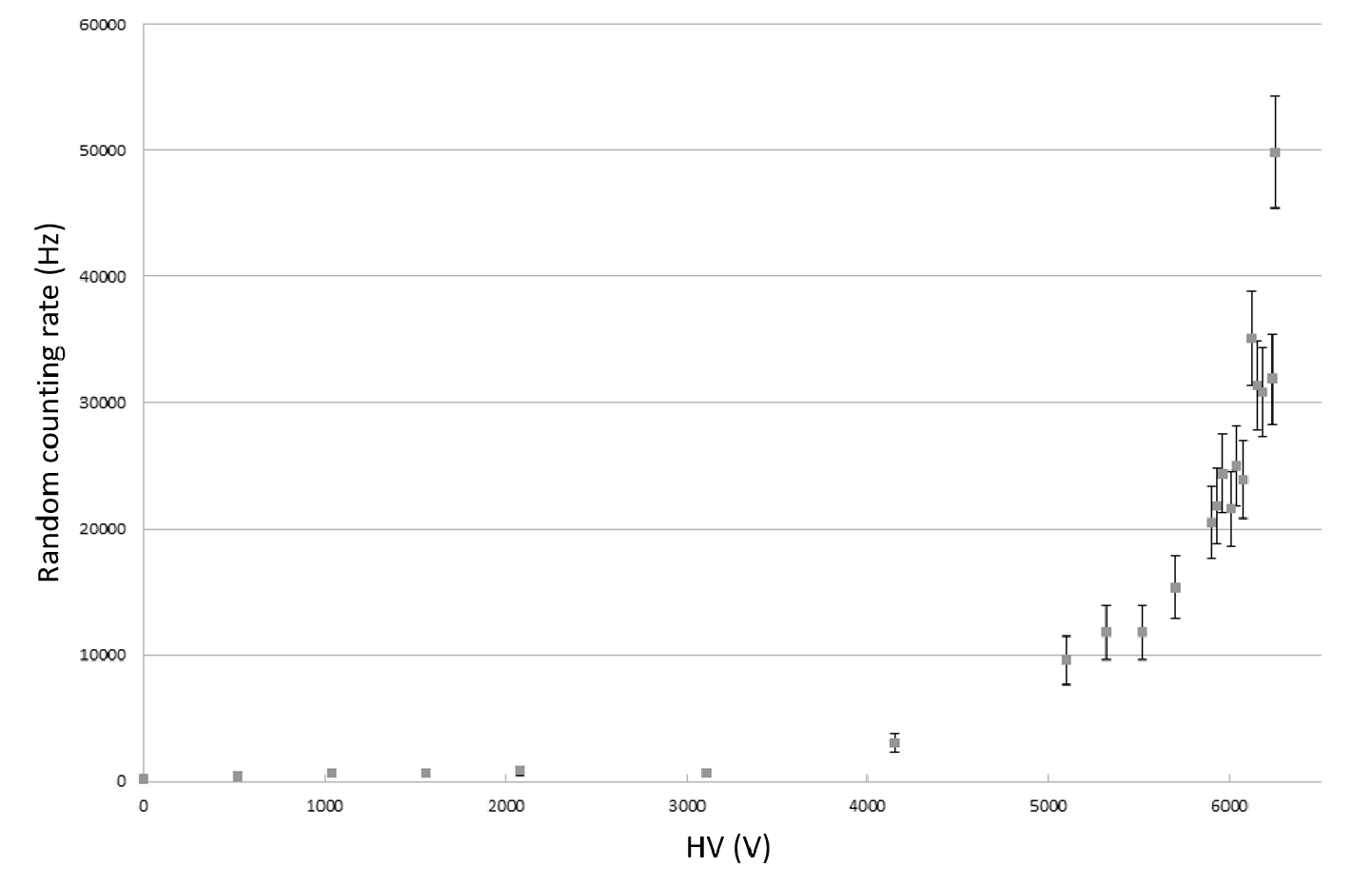}} \\
\caption{Prototype 2 results}
\label{fig:Prototype2}
\end{figure}

The characterization of the second prototype was carried out at INFN laboratories of Rome Tor Vergata using atmospheric muons. Two scintillators have been used as the trigger reference. The trigger-time resolution has been measured during the test, resulting in $(456\pm14)\; \rm ps$.
The final results are shown in fig. \ref{fig:Prototype2}. The time resolution has been evaluated as described above: a jitter of $(1.10\pm0.09)\;\rm ns$ has been measured. The 'efficiency times acceptance' curve shows a knee point at about $6200\;\rm V$. Random counting rate has been measured by acquiring waveforms in a $10\;\mu\rm s$ time window and are showed in fig. \ref{fig:Prototype2} (d). The high random counting rate is ascribed to the roughness and the imperfection of the spacers.

\begin{figure}[h]
\centering
\subfloat[][\small{Ionic signals}]
{\includegraphics[trim=0.cm 0.cm 0.cm 0.cm, clip=true,width=0.45\textwidth]{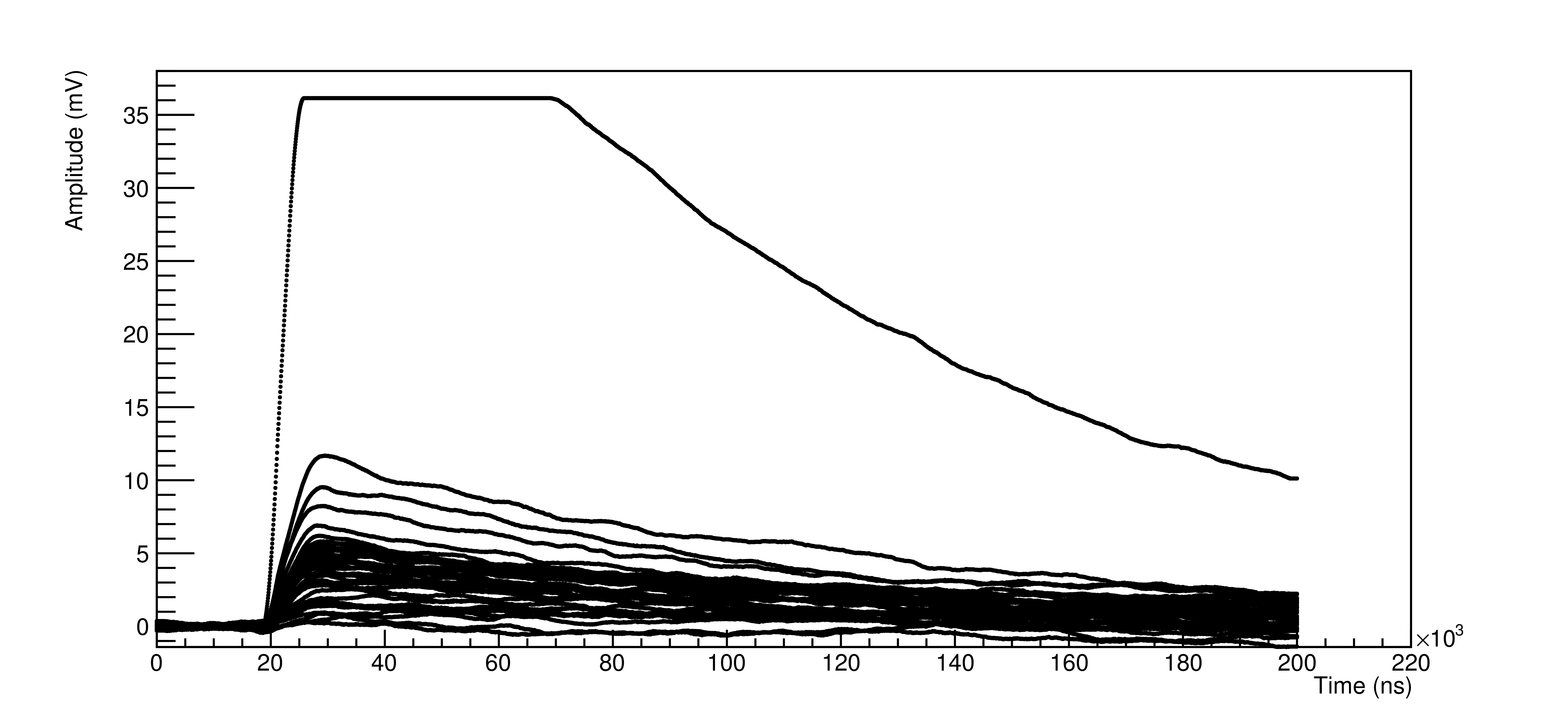}} \quad
\subfloat[][\small{Electronic signals}]
{\includegraphics[trim=0.cm 0.cm 0.cm 0.cm, clip=true,width=0.45\textwidth]{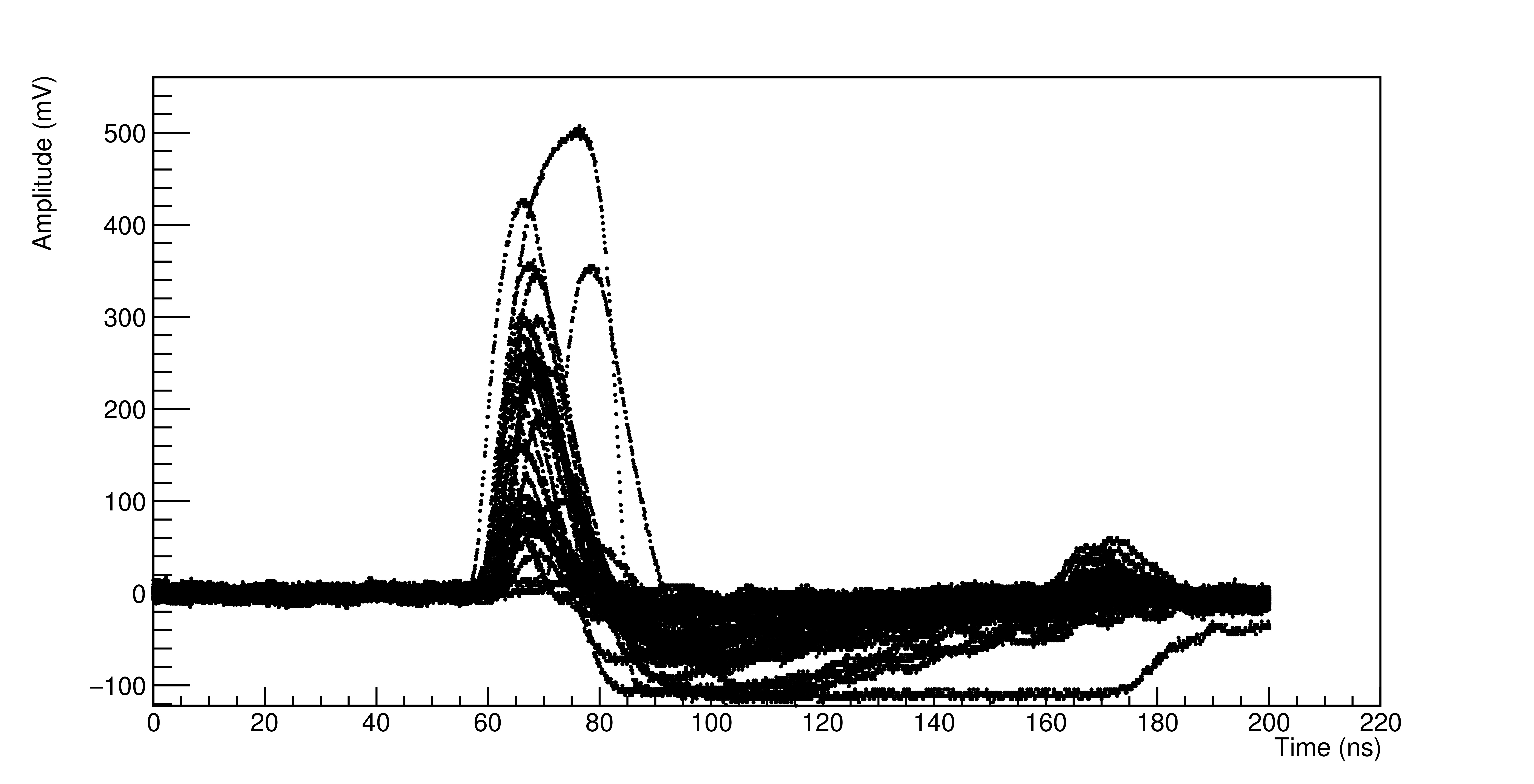}} \\
\caption{Prototype 3 pulses shaping.} 
\label{fig:Prototype3}
\end{figure}

\begin{figure}[h]
\centering
\subfloat[][\small{Time difference with respect to one scintillator detector corrected for the time-walk effect (prototype 3);}]
{\includegraphics[trim=0.cm 0.cm 0.cm 0.cm, clip=true,width=0.44\textwidth]{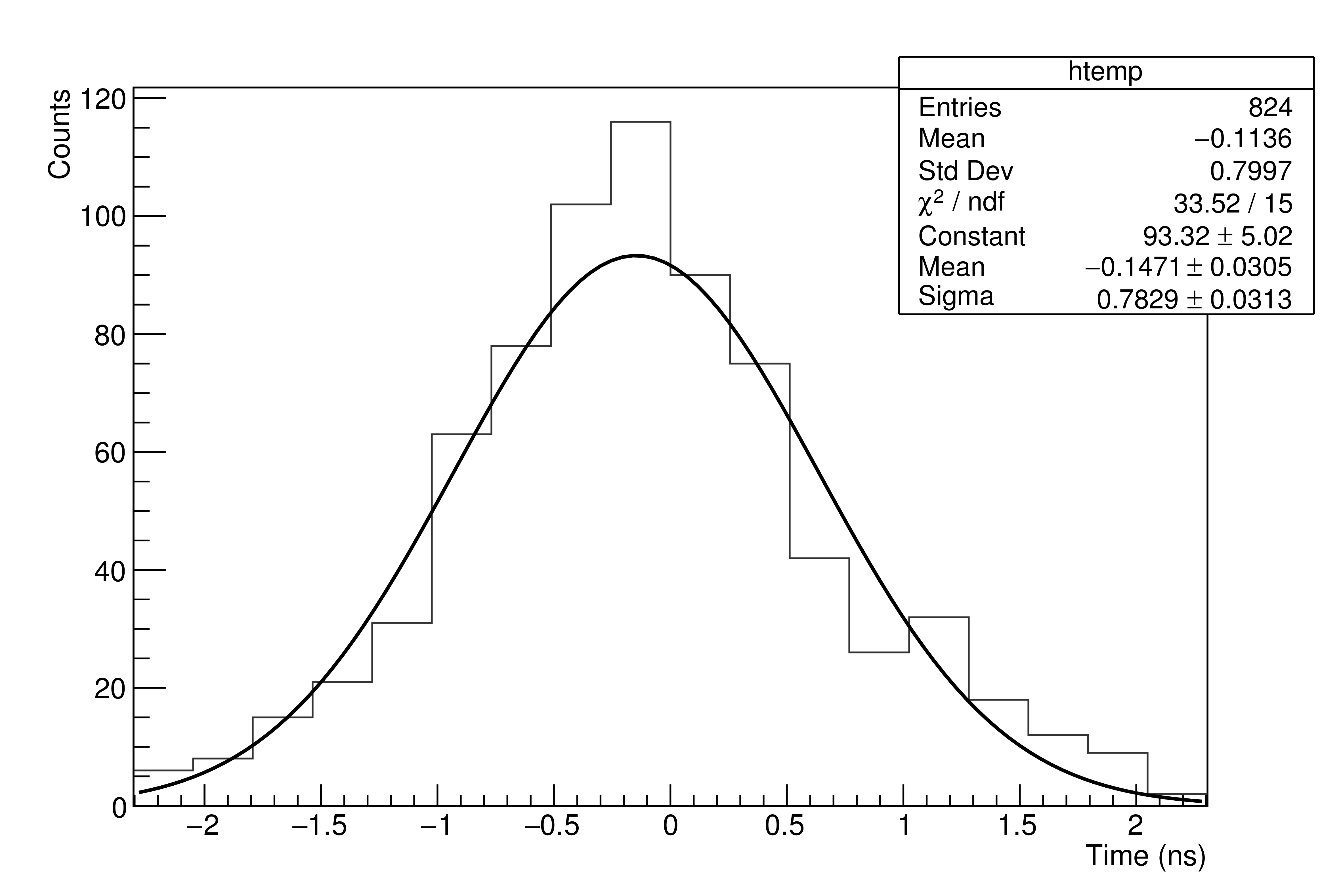}} \quad
\subfloat[][\small{'Efficiency times acceptance' ($1.3\;\rm mm$ gas gap, prototype 3);}]
{\includegraphics[trim=0.cm 0.cm 0.cm 0.cm, clip=true,width=0.5\textwidth]{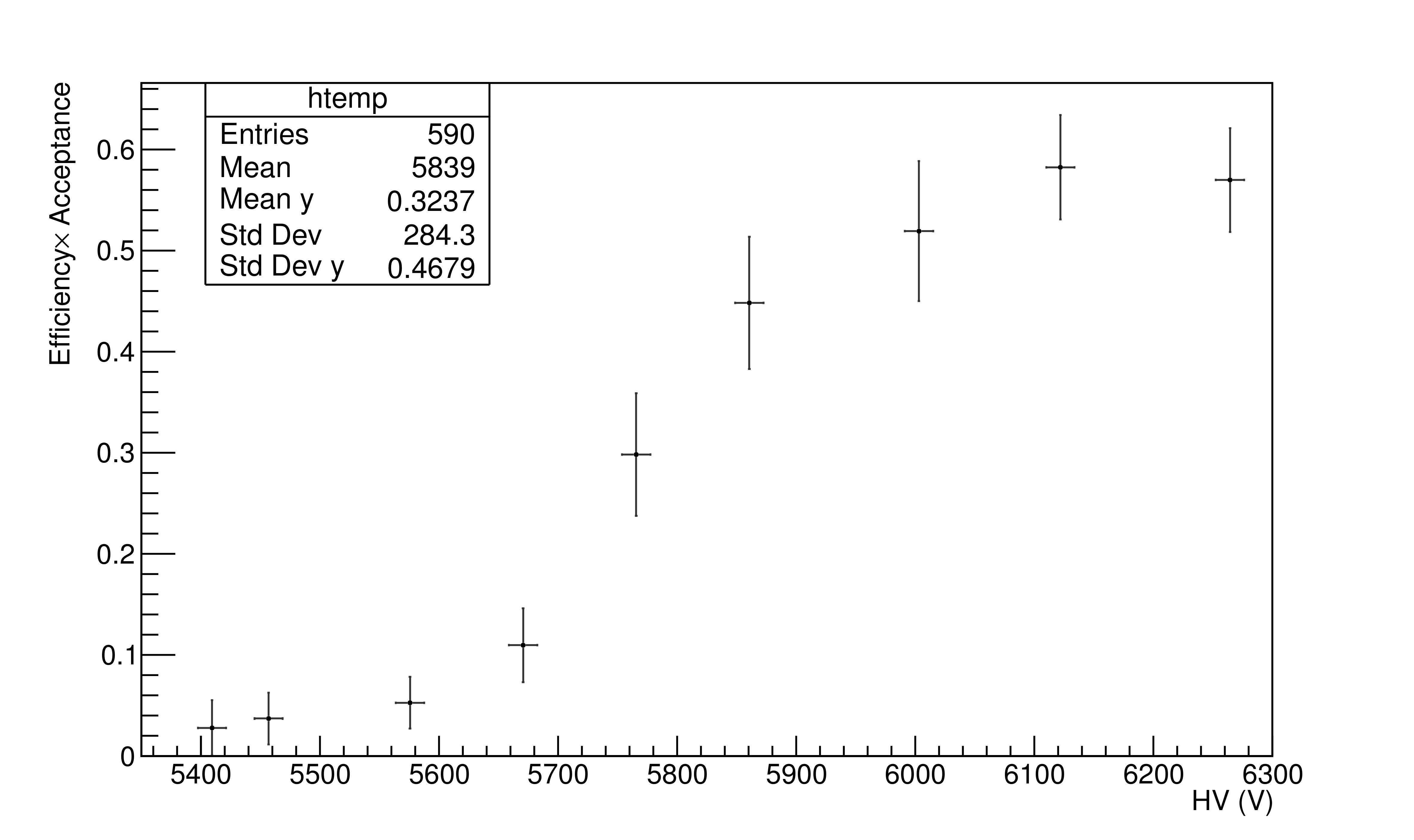}} \\
\subfloat[][\small{Random counting rate as a function of high voltage (surface$\sim4\;\rm cm^2$).}]
{\includegraphics[trim=0.cm 0.cm 0.cm 0.cm, clip=true,width=0.45\textwidth]{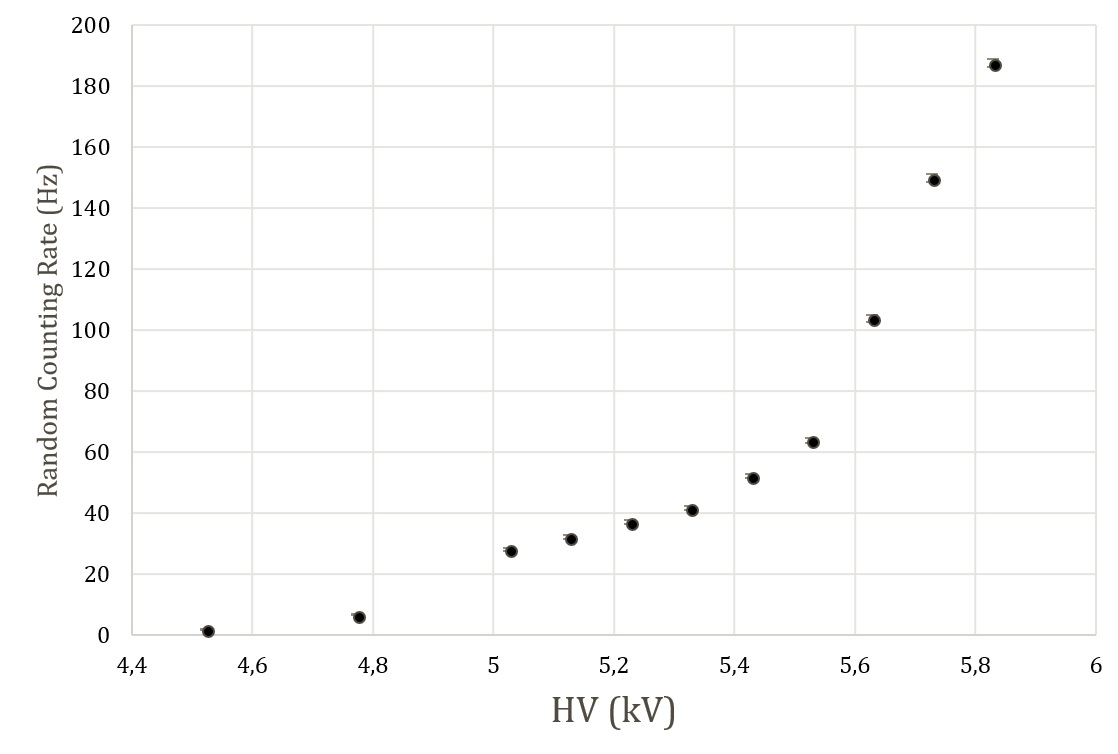}} \\
\caption{Ptototype 3 results.}
\label{fig:Prototype3/2}
\end{figure}

The characterization of the third prototype was carried out at INFN laboratories of Rome Tor Vergata using atmospheric muons. Two scintillators and one standard RPC with $2\;\rm mm$ gas gap have been used as the trigger reference. The scintillator resolution has been measured during the test, resulting in $267\; \rm ps$. The build quality of this detector made the ionic signal readout possible. In fig. \ref{fig:Prototype3} ionic and electronic signal are compared, it can be showed how the ionic signal duration exceed of four order of magnitude the electronic signal duration and there are not random counts in a such large time window. 
The final results are shown in fig.\ref{fig:Prototype3/2}. The time resolution has been evaluated as described above: a jitter of $736\;\rm ns$ has been measured. The 'efficiency times acceptance' curve shows a knee point at about $6000\;\rm V$. Random counting rate has been measured counting the signals discriminated with $30mV$ threshold in a costant time window. The results are showed in fig. \ref{fig:Prototype3/2} (d).

\newpage

\section{Conclusions}
The results obtained in these preliminary tests provide a solid foundation for the development of a new type of RPC detector. The efficiency knee as well as the time resolution are consistent with the standard RPC performances in spite of the extremly high random counting rate. Stability and low noise reached in prototype 3 allow to conduct careful studies on ionic/prompt ratio in addition to process dynamics. For an experimental confirmation of the rate-capability increase a test has been planned.
Further studies are needed to investigate the detector stability and the interactions between the gas-mixture components and the electrode surface. The behavior of the Si-GaAs subject to the voltage drop and electron flux  during the avalanche discharge is of pryority insterest.

\acknowledgments

The authors express their fond thanks to Luigi Di Stante (University of Roma "Tor Vergata") for his technical support.



\begin{thebibliography}{99}

\bibitem{a}
{R. Santonico, R. Cardarelli}, \emph{NIM} {\bf 187} (1981) 377-380.

\bibitem{b}
{R. Cardarelli, R. Santonico}, \emph{NIM} {\bf 200} (1988) 263.

\bibitem{i}
{R. Cardarelli}

\bibitem{c}
{R. Cardarelli, V. Makeev, R. Santonico}, \emph{NIM} {\bf 382} (1996) 470-474.

\bibitem{d}
{P. Camarri, R. Cardarelli, A. Di Ciaccio, R. Santonico}, \emph{NIM} {\bf 414} (1998) 317-324.

\bibitem{e}
{G. Aielli et al.}, \emph{Improving the RPC rate capability
}, \emph{Jinst} {\bf 11} (July 2016).

\bibitem{f}
\emph{ATL-COM-MUON-2017-033}, (July 7, 2017).

\bibitem{g}
{R. Cardarelli et al.}, \emph{Jinst} {10.1088/1748-0221/11/03/P03011} {IOP for SISSA Medialab} {9/2016}.

\bibitem{h}
{M. Benoit et al.}, \emph{Jinst} {10.1088/1748-0221/11/03/P03011} {IOP for SISSA Medialab} {9/2016}.

% Please avoid comments such as "For a review'', "For some examples",
% "and references therein" or move them in the text. In general,
% please leave only references in the bibliography and move all
% accessory text in footnotes.

% Also, please have only one work for each \bibitem.


\end{thebibliography}
\end{document}